**Effect of pH on structure and surface charge of $Fe_2O_3$ nanoparticles synthesized at different pH conditions and correlation to antibacterial properties**

Farzana Naushin[2†], Srishti Sen[3†], Mukul Kumar[4†], Hemang Bairagi[3†], Siddhartha Maiti[3], Jaydeep Bhattacharya[2*], Somaditya Sen[1*]

1. Department of Physics, SMART Lab, Indian Institute of Technology, Indore-453552, India

2. School of Biotechnology, Jawaharlal Nehru University, New Delhi-110067, India

3. School of Biotechnology, Jawaharlal Nehru University, New Delhi-110067, India

4. Department of MEMS, Indian Institute of Technology, Indore-453552, India

**† Equal Contribution: Please see Supplementary.**

**\*Corresponding author-** Somaditya Sen, Department of Physics, SMART Lab, Indian Institute of Technology, Indore-452001, India; orcid.org/0000-0002-6129-8344; Email: sens@iiti.ac.in

Jaydeep Bhattacharya, School of Biotechnology, Jawaharlal Nehru University, New Delhi-110067, India; orcid.org/0000-0001-7268-0867; Email: jaydeep@jnu.ac.in

**Abstract:**

pH of a solution is the ratio of H+/OH- ions. The relative ratio of these charges may affect forming bonds during a hydrothermal synthesis by influencing electronic clouds of participant ions, which can modify the structure and hence crystallinity, strain, disorder, surface termination etc. These factors may modify physical properties including the surface charge. This work uses hematite nanoparticles to correlate the structural modifications to all these properties and finally to the antibacterial properties due to the surface charge interaction of the nanoparticles and the bacterial cell walls.

Hematite, α-$Fe_2O_3$ [Sup.#1] is an n-type semiconductor(1) with a band gap corresponding to visible light energy ($E_g$ = 2.1 eV). It is non-toxic, can be easily produced, and has good corrosion resistance and antibacterial properties(2,3). Amongst numerous solution/vapor-based techniques, precipitation and hydrothermal synthesis [Sup.#2,3] are versatile routes that can generate nanoparticles of different shapes and morphology(4) due to several factors like temperature, pressure, ambient chemical conditions, precursors, and pH of the solution(5). pH being a precise measure of the relative concentration of $H^+$ (a proton in solution) and $OH^-$ ions(6), can control the electronic charge distribution of the ions participating in the reaction. Hence, it is an important factor affecting the nature and details of the forming bonds of synthesized materials thereby modifying its structure. It has been observed that different surface terminations may lead to different morphology and also electronic properties(7). Bio-related applicability such as antibacterial properties [Sup.#4,5] depend on the electrostatic forces between the surface charges of a particle and the bacterial cell charge(8). Hence, a connection between the formation conditions and the structure can be correlated and connected to the antibacterial properties: an interesting field of study. This work is an attempt to correlate the structural/electronic modifications of α-$Fe_2O_3$ nanoparticles synthesized under different pH conditions to their antibacterial ( *E. coli*) properties [Sup.#4].

Hematite nanoparticles were synthesized via a precipitation method, using Fe nitrate (salt) and ($NH_4$)OH (base), keeping the pH values of the synthesis solutions at 8, 10, and 11.52 [Sup.#1]. The three samples were thereby named S8 for pH~8, S10 for pH~10, and S11 for pH~11.52.

A Supra55 Zeiss Field Emission Scanning Electron Microscopy (FESEM) instrument(9) was used to obtain images at 15 kV with 15K magnification to study the morphology and particle size [Sup.#6]. The FESEM images (Figure 1) revealed agglomerated spherical nanoparticles of size 20~40 nm with similar morphology for the S8 and S10 samples but with smaller nanodot-like structures, of size < 5 nm, covering the entire surface. The average size of the S10 appeared to be smaller than the S8 sample. However, the S11 sample revealed much smaller particles of the size 10~30 nm. The dotlike surface morphology was present in the S11 sample. 2-D Energy Dispersive X-ray Spectroscopy (EDX) [Sup.#7] scans revealed the presence of only Fe and O and the absence of any other elements proving the phase purity of the samples.

Hydrodynamic size(10) of nanoparticles along with their Poly-Dispersity Index value (PDI)(11) were obtained from Dynamic Light Scattering (DLS) measurements(12) using Malvern Zetasizer (Nano series) [Sup.#8]. The stability and surface charge of the nanoparticles were estimated from their Zeta potential (ζ-potential)(13) using the same instrument. Manual readings for 3 measurements of 6 cycles for each nanoparticle were taken. The PDI value of the samples was

found to be 0.540±0.059 (S8), 0.386±0.026 (S10), and 0.570±0.048 (S11) [Figure 1g]. While S11 seems to be definitely polydispersive (PDI > 0.5) and S10 to be monodispersive (PDI < 0.5) in nature, the S8 sample seems to be a borderline case with a tendency of polydispersive nature. The hydrodynamic size of the nanoparticles in solution seems to be least for the S10 sample ~186.6±2.5 d.nm, with a much bigger S8 ~467.1±43.3 d.nm and comparatively larger S11 ~265.9±15.4 d.nm [Figure 1 [Sup.Table#1]. This contrast in the dispersive nature and hydrodynamic size of the particles can be a consequence of the surface charge of the nanoparticles. The sign of the $\zeta$-potential ensures the nature of the surface charge of nanoparticles. The $\zeta$-potential value recorded for the three samples were -4.85±0.90 mV (S8), +11.10±0.10mV (S10) and +22.10±1.00 mV (S11) [Figure ]. Hence, one can observe that the nature of the surface charge has modified from a negative to a positive charge from S8 to S10 and continues to increase in magnitude in S11 [Sup.Table#1]. The magnitude of the $\zeta$-potential also depicts the stability of a nanoparticle and hence S11 seems to be the most stable followed by S10 and S8.

X-Ray Diffraction (XRD) studies(14) [Sup.#9] [using a Bruker D2 Phaser x-ray diffractometer] revealed a crystalline trigonal Hematite ***α***-Fe$_2$O$_3$ *[space group R-3c]* structure for all the samples [Figure 2]. Rietveld refinement(15) of the XRD data was performed using GSAS software to extract the structural details. The lattice parameters a=b were found to be maximum for S8 (~5.02925 Å) than a much smaller S10 (~4.97272 Å) and S11 (~ 4.97302 Å). The lattice parameter 'c' revealed a minimum for S10 (~13.57733 Å) with a comparable but larger S11 (~ 13.58262 Å) and a much larger S8 (~13.73285 Å) [Figure 1(a)] [Sup.Table#2]. Hence, the unit cell volume of the S10 sample is least followed by S11, compared to a much larger S8. Therefore, the density is highest for S10 sample. It appears that the structure is affected by the pH of the preparation solvent. Such subtle changes bear its roots in changing bondlengths and bondangles which are dependent on the properties of the electronic hybridization during the bonding process. The refined Crystallographic Index Files (CIF)(16) were analyzed using Mercury software. Bond lengths and angles were extracted [Sup.#10]. The structure of $Fe_2O_3$ is an assembly of $FeO_6$ octahedra with an O-lattice and a Fe-lattice. The O-lattice consists of O-planes with each O-atom being connected to neighboring six O-atoms, thereby creating four longer bonds $(O-O)_l$ and two shorter bonds $(O-O)_s$ [Figure 1(c,d)]. The Fe-atoms are in a staggered plane in between two O-planes. The Fe-Fe bonds within the same plane, $(Fe-Fe)_p$, are 2.944 Å (S8) and 2.911 Å (S10 and S11). The Fe-Fe interplanar bond, $(Fe-Fe)_i$, between subsequent Fe-layers are shorter than $(Fe-Fe)_p$. $(Fe-Fe)_i$ decreases significantly from 2.777 Å (S8) to 2.745 Å (S10) and 2.746 Å (S11). The $O_s$-Fe-$O_l$ bond angle in the $FeO_6$ octahedron was ~89.09° for S8 and 89.1° for S11 but is much smaller for S10 ~87.24°. However, the other two angles, i.e. $O_l$-Fe-$O_l$ (~82.28°) and Os-Fe-Os (~99.95°) show nominal changes and can be considered unaffected by the pH. The $O_s$-Fe-$O_l$ angle is least for the S10 sample saying that the $FeO_6$ octahedra is most distorted for this sample and the hexacoordinated Fe is most off-centered in the $O_6$-cage. Two different sets of three Fe-O bonds are observed with three shorter Fe-$O_s$ bonds in one plane and three longer Fe-$O_l$ bonds in the adjacent O-plane. The distortion can be also revealed from these bond lengths. The values of Fe-$O_l$ are 2.111 Å for both S10 and S11 but is longer for the S8 sample ~2.135 Å, while Fe-$O_s$ keep on reducing from 1.928 Å in S10, to 1.908 Å for S10 and 1.906 Å for S11 [Table Sup. 3]. These subtle changes in the bondlengths and bondangles not only distort the $FeO_6$ octahedra but also must have consequences on the surface terminations of different crystallographic planes. The XRD data was investigated for this purpose from a different light, in which a comparative study was performed on the subtle changes in the peak intensity of different lattice planes. The peak intensities of the different lattice planes were normalized to the strongest intensity and the relative strength of different peaks were analyzed. It was observed that these relative intensities almost remained invariant for most of the planes when considering S8 and S10, except for (2 1 4) and (1 0 10) planes [Figure 1(b)]. All the planes showed changes for S11. The relative intensity of the (2 1 4) plane decreased continuously from S8 to S11 while that of (1 0 10) increased. Each of these planes was investigated in VESTA software and it was found that the (2 1 4) plane terminates with negative O-atoms, while the (1 0 10) plane terminates with both O and Fe ions. Hence, the reduction of the relative intensity of the (2 1 4) plane with increasing pH may be a prime cause behind reduction of negative surface charge of the nanoparticles. The d-spacings of the lattice planes were also found to vary in between samples. Thus, it is evident from the above structural analysis that the pH does modify the hybridization of the participant ions and results in modification of the bondlengths and bondangles, thereby modifying the structure and preferential termination of the nanoparticles. These factors in turn modify the surface charge of the materials.

Strain and crystallite size were calculated using Scherrer(17) and Williamson-Hall equation(18) [Sup.#11]. The strain in the S8 sample is 0.0042, which nominally decreases for S10 to 0.0041 and thereafter increases noticeably to 0.0053 for S11. Such changes can now be correlated to the fact that the most change in the crystallographic planes is observed for the S11 sample. As a consequence of strain, the average crystallite size increased nominally from ~30.17 nm for S8, to 30.48 nm for S10 and thereafter reduced to a much smaller size of ~23.87 nm for S11 [Sup.Table#2]. It is noteworthy that despite the largest crystallite size of the S10 sample, the hydrothermal size of the same is the minimum among the three samples. This may be the consequence of the changing nature and magnitude of the surface charge which seems to be correlated to the modifying bondlengths of the materials.

Fe$_2$O$_3$ has been reported in the literature to have both a direct and indirect bandgap. The direct and indirect bandgap ($E_g$)(19) and lattice disorder ($E_U$, Urbach energy)(20) [Sup.#12] were studied using the optical reflectance, $R$, of the samples using a UV-Vis (Research India Spectrophotometer). In this study, the direct bandgap, $E_{g\text{-}direct}$ was found to continuously increase from 2.127 eV (S8), 2.173 eV (S10) to 2.198 eV (S11). The indirect bandgap, $E_{g\text{-}indirect}$ was found to decrease from 1.917 eV (S8) to 1.873 eV (S10) and 1.807eV (S11). However, the errors were high for both $E_{g\text{-}direct}$ and $E_{g\text{-}indirect}$ to add any meaningful change in $E_g$ [Figure 1(e)]. Hence, the probable explanation of differences of Reactive Oxygen Species (ROS) [Sup#5] behind changes in antibacterial properties can be disqualified from this observation. However, the Urbach energy reduces from 21.803 meV in S8 to 11.86 meV in S10 and thereafter increases drastically to 56.14 meV in S11 [Figure 1(g)]. Such changes are also consistent with the changes in structural studies.

Observing such changes in the structural and electronic properties of the materials, it was obvious to check the antibacterial properties of these materials as one of the routes to account for the antibacterial properties of the nanoparticles is via an electrostatic interaction between the opposite charges of the cell walls and the nanoparticles, which can lead to cell wall rupture(21) . To study the effect of pH-controlled Fe$_2$O$_3$ samples on the viability of Gram-negative bacteria *E. coli* DH5α ATCC 25922 [Sup.#4] all the nanoparticles were first sterilized using ethanol. The bacterial cells were revived, and a master plate was prepared from it. A single colony was picked and inoculated in Luria Bertani broth (LB). This primary culture was left for overnight incubation at 37°C, 200 rpm. From the primary culture, a secondary culture was prepared to have 0.5 optical density. The bacterial culture was then exposed to various concentrations of nanoparticles and LB media which served as a negative control. Nanoparticles alone served as particle control. The plates were incubated at 37ºC for 18 hours then OD at 600nm was measured and the percentage of cell viability was calculated(22). All experiments were triplicated. 50% of cell viability (IC50) was observed at a concentration of 738.91±116.7 µg/ml for S11, 418.99±7.3 µg/ml for S10, and 766.05±70.19 µg/ml for S8 respectively [Figure 1(a)]. In this study, S10 revealed the lowest IC 50 value, i.e., maximum antibacterial properties. The negative charge of the majority of S8 samples and the high positive charges of the S11 sample seems to be adverse to the nanoparticle-cell wall interaction, while the S10 sample with a moderate charge and a smaller hydrodynamic size seem to be the most effective antibacterial material.

Hence, the effect of different pH of the ambient solution during the synthesis of the $Fe_2O_3$ nanoparticles is noteworthy. The different ratios of $H^+$ and $OH^-$ ions modify the electronic clouds of the cations and anions and thereby modify the bonding. From the structure aspect, it is observed that the unit cell volume decreases, i.e., the density increases, as the pH is raised from 8 to higher values. This is due to the contracting bondlengths of the $FeO_6$ octahedra. These changes have an effect on the lattice strain which is minimum for the S10 sample. As a result, the crystallite size is maximum for the S10 sample. This results in the crystalline disorder being the least for the S10 sample. The changes in the bondlengths are such that one of the angles $O_l$-Fe-$O_s$ is drastically less for the S10 sample, which may be correlated to the monodispersive nature of the sample. These changes lead to the modifications of the offcentering of the Fe atom in the $FeO_6$ octahedra. It was observed that the surface charge of the nanoparticles changed from negative to positive when the pH changed from 8 to 10 and kept on increasing with increasing pH. The changes in the bondlengths and bondangles bring subtle changes to the surface charge and are close to a neutral value in between a pH of 8-10, where the angle is the least. A least positive charge on the surface may be correlated to the monodispersive nature of the nanoparticles and allows the same to obtain the least value for the hydrodynamic size in spite of the maximum crystallite size of the same.

Such a property allows more particles to be associated with the *E.Coli* bacteria and thereby form electrostatic interactions with the same. The negative charge of S8 should repel the bacterial surface while the polydispersive S11 in spite of its higher charge has a lesser chance of interaction due to the agglomerated nature.

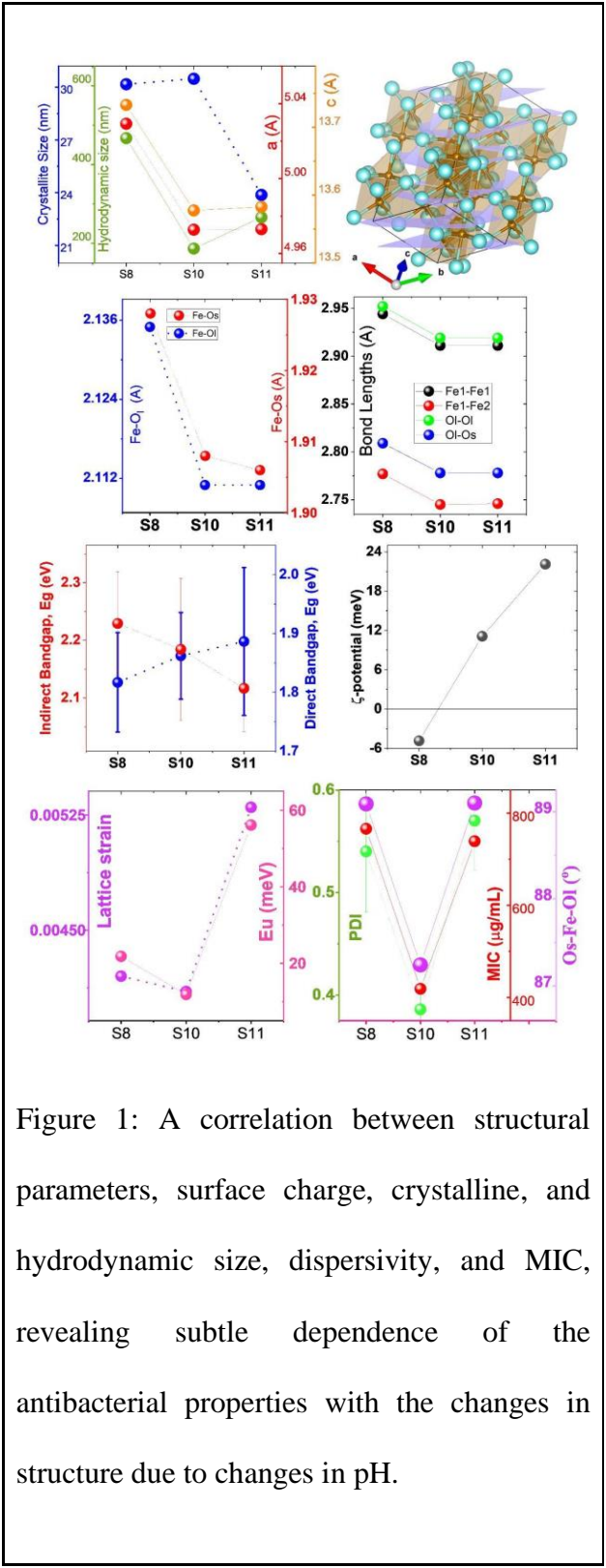

Figure 1: A correlation between structural parameters, surface charge, crystalline, and hydrodynamic size, dispersivity, and MIC, revealing subtle dependence of the antibacterial properties with the changes in structure due to changes in pH.

The combination of a smaller hydrodynamic size lets more nanoparticles access the bacterial surface (supports greater mobility of the particles) and the smaller PDI which ensures shorter separation of the nanoparticles, enables the negative bacterial surface to interact with the less positive S10 nanoparticles than the negative S8 and the more positive S11 nanoparticles.

Hence, this work reveals the importance of the subtle changes in the chemical environment during synthesis that may lead to subtle changes in bond lengths of a particular composition and thereby modify important factors such as surface charge and dispersive nature of the particles, further affecting the responses towards biological entities like bacteria.

This is a new observation that needs attention from all drug manufacturers dealing with nanomedicine and is most probably the gateway to a new physics/chemistry related to biology.

**Conclusion:**

The pH of the ambiance of hydrothermal synthesis of hematite nanoparticles affects the bonding nature of the Fe-O bonds in the $FeO_6$ building blocks of the $\alpha$-$Fe_2O_3$ structure. The bonds shorten as pH is increased from 8 to contract the unit cell and increase the density. Hence, the lattice strain is modified (least for S10) thereby modifying the crystallite size (largest for S10). A Fe-off-centering is observed which seems to increase the surface charge from negative to positive and higher. The nominal positive charge of S10 samples is most likely the reason for the smaller hydrodynamic size and hence provides better movability on the bacteria surface. This enhances the chance of bonding between negatively charged bacterial surfaces and positively charged S10 nanoparticles. Thus, the chances of bacterial cell rupture increases.

**Acknowledgment:**

Dr. S. Sen is thankful to the Department of Science and Technology (DST), Govt. of India, for providing funding through grant no: DST/TDT/AMT/2017/200. The authors are grateful for the financial support provided by the Council of Scientific and Industrial Research (CSIR), Government of India. The authors would like to acknowledge the Sophisticated Instrument Center (SIC), IIT Indore for providing the FESEM facility.

The authors declare no conflict of interest.

**References:**

1. Ali A, Zafar H, Zia M, ul Haq I, Phull AR, Ali JS, et al. Synthesis, characterization, applications, and challenges of iron oxide nanoparticles. Nanotechnol Sci Appl. 2016 Aug;Volume 9:49–67.
2. Sirelkhatim A, Mahmud S, Seeni A, Kaus NHM, Ann LC, Bakhori SKM, et al. Review on Zinc Oxide Nanoparticles: Antibacterial Activity and Toxicity Mechanism. Nanomicro Lett. 2015 Jul 19;7(3):219–42.
3. Vihodceva S, Šutka A, Sihtmäe M, Rosenberg M, Otsus M, Kurvet I, et al. Antibacterial Activity of Positively and Negatively Charged Hematite ($\alpha$-$Fe_2O_3$) Nanoparticles to Escherichia coli, Staphylococcus aureus and Vibrio fischeri. Nanomaterials. 2021 Mar 8;11(3):652.
4. Farahmandjou M. Synthesis and characterization of $\alpha$-$Fe_2O_3$ nanoparticles by simple co-precipitation method. Research Letters in Physical Chemistry. 2015;
5. Szterner P, Biernat M. The Synthesis of Hydroxyapatite by Hydrothermal Process with Calcium Lactate Pentahydrate: The Effect of Reagent Concentrations, pH, Temperature, and Pressure. Bioinorg Chem Appl. 2022 Mar 25;2022:1–13.
6. Buxton G v., Greenstock CL, Helman WP, Ross AB. Critical Review of rate constants for reactions of hydrated electrons, hydrogen atoms and hydroxyl radicals ($\cdot OH/\cdot O^-$ in Aqueous Solution. J Phys Chem Ref Data. 1988 Apr;17(2):513–886.
7. Liu N, Chen X, Zhang J, Schwank JW. A review on $TiO_2$-based nanotubes synthesized via hydrothermal method: Formation mechanism, structure modification, and photocatalytic applications. Catal Today. 2014 Apr;225:34–51.
8. Mendes CR, Dilarri G, Forsan CF, Sapata V de MR, Lopes PRM, de Moraes PB, et al. Antibacterial action and target mechanisms of zinc oxide nanoparticles against bacterial pathogens. Sci Rep. 2022 Feb 16;12(1):2658.
9. Wang D, Xie T, Li Y. Nanocrystals: Solution-based synthesis and applications as nanocatalysts. Nano Res. 2009 Jan 5;2(1):30–46.

10. Hackley VA, Clogston JD. Measuring the Hydrodynamic Size of Nanoparticles in Aqueous Media Using Batch-Mode Dynamic Light Scattering. In 2011. p. 35–52.
11. Nobbmann U, Morfesis A. Light scattering and nanoparticles. Materials Today. 2009 May;12(5):52–4.
12. Babick F. Dynamic light scattering (DLS). In: Characterization of Nanoparticles. Elsevier; 2020. p. 137–72.
13. Chen LC, Kung SK, Chen HH, Lin SB. Evaluation of zeta potential difference as an indicator for antibacterial strength of low molecular weight chitosan. Carbohydr Polym. 2010 Oct;82(3):913–9.
14. Bunaciu AA, Udriştioiu E gabriela, Aboul-Enein HY. X-Ray Diffraction: Instrumentation and Applications. Crit Rev Anal Chem. 2015 Oct 2;45(4):289–99.
15. Sakata M, Cooper MJ. An analysis of the Rietveld refinement method. J Appl Crystallogr. 1979 Dec 1;12(6):554–63.
16. Hall SR, Allen FH, Brown ID. The crystallographic information file (CIF): a new standard archive file for crystallography. Acta Crystallogr A. 1991 Nov 1;47(6):655–85.
17. Muniz FTL, Miranda MAR, Morilla dos Santos C, Sasaki JM. The Scherrer equation and the dynamical theory of X-ray diffraction. Acta Crystallogr A Found Adv. 2016 May 1;72(3):385–90.
18. Mustapha S, Ndamitso MM, Abdulkareem AS, Tijani JO, Shuaib DT, Mohammed AK, et al. Comparative study of crystallite size using Williamson-Hall and Debye-Scherrer plots for ZnO nanoparticles. Advances in Natural Sciences: Nanoscience and Nanotechnology. 2019 Nov 14;10(4):045013.
19. Yuan LD, Deng HX, Li SS, Wei SH, Luo JW. Unified theory of direct or indirect band-gap nature of conventional semiconductors. Phys Rev B. 2018 Dec 26;98(24):245203.
20. Ugur E, Ledinský M, Allen TG, Holovský J, Vlk A, de Wolf S. Life on the Urbach Edge. J Phys Chem Lett. 2022 Aug 25;13(33):7702–11.
21. Redza-Dutordoir M, Averill-Bates DA. Activation of apoptosis signalling pathways by reactive oxygen species. Biochimica et Biophysica Acta (BBA) - Molecular Cell Research. 2016 Dec;1863(12):2977–92.
22. Sarkar S, Thapa R, Naushin F, Gupta S, Bhar B, De R, et al. Antibiotic-Loaded Smart Platelet: A Highly Effective Invisible Mode of Killing Both Antibiotic-Sensitive and -Resistant Bacteria. ACS Omega. 2022 Jul 19;7(28):24102–10.

**Supplementary Information:** *"Effect of pH on structure and surface charge of Fe2O3 nanoparticles synthesized at different pH conditions and correlation to antibacterial properties"*

Farzana Naushin[2†], Srishti Sen[3†], Mukul Kumar[4†], Hemang Bairagi[3†], Siddhartha Maiti[3], Jaydeep Bhattacharya[2*], Somaditya Sen[1*]

1. Department of Physics, SMART Lab, Indian Institute of Technology, Indore-453552, India

2. School of Biotechnology, Jawaharlal Nehru University, New Delhi-110067, India

3. School of Biotechnology, Jawaharlal Nehru University, New Delhi-110067, India

4. Department of MEMS, Indian Institute of Technology, Indore-453552, India

## Justification of contribution and role of each author in the draft:

### Farzana Naushin[†] (PhD scholar, JNU):

MIC data (operator), MIC analysis (analyzer), DLS data, DLS analysis (analyzer), Writing (analyzer), Overall analysis (analyzer)

### Srishti Sen[†] (Undergraduate student BTech VIT Bhopal):

Sample Prep (operator), XRD data (operator), Mercury analysis (analyzer), UV-vis data (operator), FESEM analysis (analyzer), Preliminary Writing (analyzer), Figures (analyzer), MIC graphical analysis (analyzer), Final editing (analyzer)

### Mukul Kumar[†] (Ph.D. scholar, IIT Indore):

Sample sintering (operator), XRD data (operator), Rietveld analysis (analyzer), UV-vis data (operator), UV-vis analysis (analyzer), FESEM data (operator), EDX data (operator)

**Hemang Bairagi[†] (Undergraduate student BTech VIT Bhopal):**

Preliminary Writing (analyzer), Mercury analysis (analyzer), FESEM analysis (analyzer), Figures (analyzer), MIC graphical analysis (analyzer), Final editing (analyzer)

**Siddhartha Maiti (Supervisor, VIT Bhopal):**

Overall analysis and discussion

**Jaydeep Bhattacharya* (Supervisor JNU):**

MIC analysis (analyzer), DLS analysis (analyzer), Writing and editing (analyzer)

**Somaditya Sen* (Supervisor and PI, IIT Indore):**

Problem ideation, Execution, Synthesis, Structural analysis, XRD analysis (analyzer), UV-vis analysis (analyzer), FESEM analysis (analyzer), Figures (analyzer), Draft editing (analyzer), Final editing (analyzer)

**Supporting Information:**

- ❖ **Sup.#1:**

**Introduction to Hematite**

Nanoparticles of simple oxides like ZnO, TiO$_2$, Fe$_2$O$_3$, etc(1–4), play essential roles in various fields amongst which antibacterial properties are one of the leading interests (5). Materials can be modified at the atomic level and introduce modifications in the electronic and optical properties. This enables them to be applicable in a range of new applications (6). Amongst a wide range of nanomaterials, simple oxide nanoparticles are robust and bear steady properties. Fe-oxides have been thoroughly studied in the field of bio-applications (7,8). Depending on the different oxidation states of Fe (9–11) the literature is rich with reports on hematite ($\alpha$-$Fe_2O_3$), maghemite ($Fe_3O_4$), and magnetite ($\gamma$-$Fe_3O_4$) having different crystal structures (12). Amongst all these iron oxides hematite is probably the most studied composition in literature. It is the most important iron ore as ~70% of its content is Fe. Also, this is available in abundance. Due to its red color, and a similarity to the color of blood (Greek ~ Haima; Latin ~ Haema), the ore gets its name Haematite, or simply hematite.

❖ **Sup.#2:**

**Importance of solution-based Precipitation and Hydrothermal Process:**

Amongst numerous solution/vapor-based techniques, the hydrothermal method is a trusted and robust method capable of generating nanoparticles of different shapes and morphology (13). This method uses an aqueous medium where individual ions react and form bonds in a vessel that is generally closed. The closed environment is capable of creating an environment that can attain a high temperature and high pressure just by raising the temperature to a manageable temperature thereby creating an internal pressure (vapor pressure). Another method using the solvent route is the precipitation method which is most probably the simplest method of making iron oxide nanoparticles. The process precipitates Fe2+ and Fe3+ ions upon treatment by a base. The ions form bonds with the anions and thereby form the desired oxides. The shape and morphology depend on several factors like temperature, pressure, ambient chemical conditions, precursors, and pH of the solution (14). The pH is an important factor controlling the size and shape due to the fact that the relative concentration of $H^+$ (a proton in solution) and $OH^-$ ions can play a decisive role in the electronic charge distribution of the component cations and anions and hence determine the bonds, thereby the structure of the forming material. These may be the source of different surface terminations which in turn can be responsible for different sizes and morphology. Therefore, these are also responsible for changes in electronic properties, surface charge, etc., and hence, the applicability of the materials. Among many, electronic and bio-related applications are two major branches. A connection between the two is an interesting field.

### ❖ Sup.#3:

**Synthesis Details**

A solution-based precipitation method was used for the synthesis of α-$Fe_2O_3$ nanoparticles using $Fe(NO_3)_3 \cdot 9H_2O$ (Alfa Aesar ACS 98-101%). AR-grade $NH_4OH$ (SDFCL 25% Specific Gravity 0.91) was used as a precipitating agent. The samples were prepared with three pH values and thereby named S8 for pH~8, S10 for pH~10, and S11 for pH~11.52. For all samples, 12 g of $Fe(NO_3)_3$ was dissolved in 100 ml double distilled water and uniformly stirred on a magnetic stirrer at RPM 250 for 45 minutes. $NH_4OH$ was added dropwise to obtain the desired pH values while stirring. Precipitation happened at room temperature. Maximum yield was ensured by heating the solutions at ~60°C. The solutions were dried at 60°C to provide the dried powders which were calcined thereafter at 450°C in a muffle furnace for 6 hours.

### ❖ Sup.#4:

**Gram-positive and Gram-negative bacterial strains**

Gram-positive and Gram-negative bacterial strains are responsible for human diseases, with the gram-negative bacteria having higher inhibitory properties due to their complex structure (15). Hence, this study is confined to the effects of $Fe_2O_3$ nanoparticles on a gram-negative bacterial strain namely *E. coli* DH5α. Adsorption on the bacterial surface, formation of different intermediates, and electrostatic interactions may be some of the pathways of annihilation of *E. coli* in presence of oxide nanoparticles. A thin layer of peptidoglycan is present between the two membranes on the cell surface of *E. coli*. This layer is responsible for antimicrobial resistance. More negative charges are produced due to the dissociation of carboxyl groups in the membranes. The source of the negative charges on the cell surface is the lipoteichoic acid in the membrane and the teichoic acid in the peptidoglycan layer. The surface charges can be studied by ζ-potential measurements. The opposite charges between the negatively charged cell membranes and the positive charges of the surface of the nanoparticles can result in electrostatic attractive forces. The electrostatic gradient between these two can be the source of damage to the cell membranes. This aspect of the surface charge properties has been investigated in detail to correlate the structural properties with the growth inhibition of *E. coli* DH5α (16,17).

**Sup.#5:**

**Antibacterial properties**

Reactive Oxygen Species (ROS) mediated bactericidal effect is a well-known phenomenon for iron oxide nanoparticles ((2,18,19). The electronic properties of these nanoparticles play a major role in ROS generation and thereby affect their antibacterial activities. Mostly, the higher the ROS, the more antibacterial property. Amongst various other inhibition pathways for *E. coli*, the electrostatic interaction seems to be a probable mechanism in the case of the $Fe_2O_3$ nanoparticles. From the ζ-potential studies, one can observe the positive values of the surface charge for the S10 and S11 samples. The opposite charges between the negatively charged cell membranes and the positive charges of the surface of the S10 and S11 samples result in attractive electrostatic forces. The electrostatic gradient between these two units damages the cell membranes. Hence, one can justify the response of the S10 and S11 samples toward the annihilation of the *E. coli* DH5α.

The $Fe_2O_3$ samples had a moderate band gap of ~2.13 eV. This energy corresponds to a greenish-yellow light. An illumination of white light consisting of more energetic components of light such as green, blue, violet, or UV can hence promote an electron in the valence band (VB) to the conduction band (CB) overcoming the band gap of the $Fe_2O_3$ material, thereby leaving a hole in the VB. This is a light-assisted mechanism of the formation of an electron–hole pair in the material. The hole generated in the valence band (VB) gets a strong oxidizing nature. Thus, there is a light-assisted creation of oxidizing sites. These can oxidize $H_2O$ or $OH^-$ to produce strong oxidizing species. A redox chain reaction follows. ROS are formed, e.g. hydroxyl radical ($\cdot OH$), hydroperoxide radical ($\cdot H_2O_2^-$), and superoxide radical anion ($O_2 \cdot^-$). These are the pathways of bactericidal action. Thus ROS generation from $Fe_2O_3$ samples can induce oxidative stress in the bacterial cell. This may lead to protein synthesis inhibition and DNA replication. A destabilization mechanism of the charges present in the cytoplasmic membrane may happen due to the electronic excitation and rupture of the membrane.

It is generally observed that smaller nanoparticles can move freely over the membrane surface, and thus the electrostatic interactions are better. Thus, the smaller hydrodynamic size, $d$, of the S10 sample may be a possible reason for the better movement of these particles, and the moderately positive surface charge of ~11.1mV is most likely the reason for a higher probability of membrane rupture.

Thus the enhanced antibacterial property of S10 seems to be correlated to the positive surface charge and the smaller hydrodynamic size of the particles, which is a consequence of the influence of alkalinity on the structural properties. Further, the lowest Urbach Energy hints at a less disordered lattice which enables the annihilation of the disorder-related tail states of the band edges, thereby raising the indirect band gap. This less disordered lattice enables easier phonon-assisted e-h pair generation upon illumination of white light which in turn leads to higher Reactive Oxygen Species formation.

## ❖ Sup.#6:

**FESEM studies**

The morphology and particle size of the synthesized particle was studied using Field Emission Scanning Electron Microscopy (FESEM). A Supra55 Zeiss FE-SEM instrument was used for the analysis. The instrument could be operated in a voltage range 0.02-30 kV, however, most of the images were obtained at 15 kV for a magnification of 15K.

*Size and Surface morphology*

The FESEM images (Figure 1) revealed agglomerated spherical nanoparticles of size 20~40 nm for all the samples with further smaller nanodot-like structures, of size < 5 nm, covering the entire surface. A very similar morphology is observed for S8 and S10 samples. The average size of the S10 seems to be a bit smaller than the S8 sample. However, the S11 sample actually revealed much smaller particles of the size 10~30 nm. The dotlike surface morphology was present in the S11 sample but was less detectable due to the charging nature of the sample. This revealed a lesser conductive nature of the S11 sample.

*Observations*

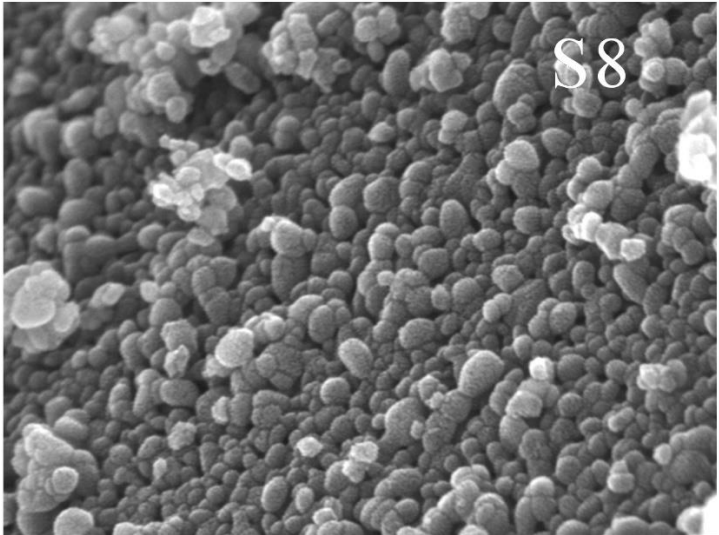
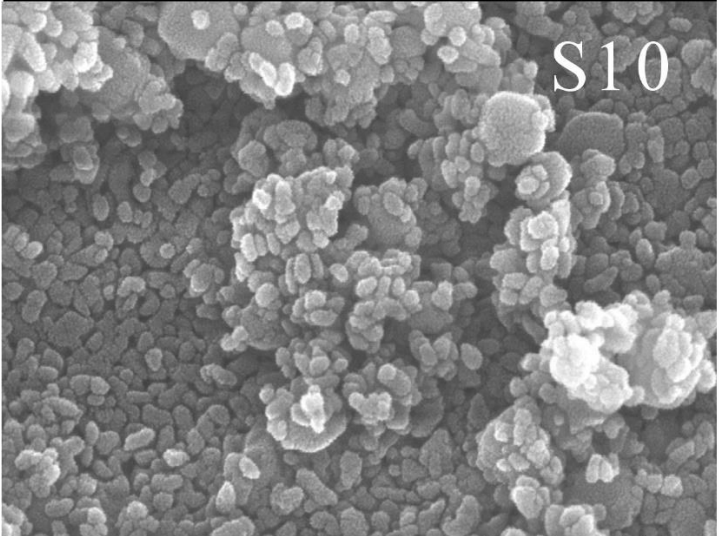
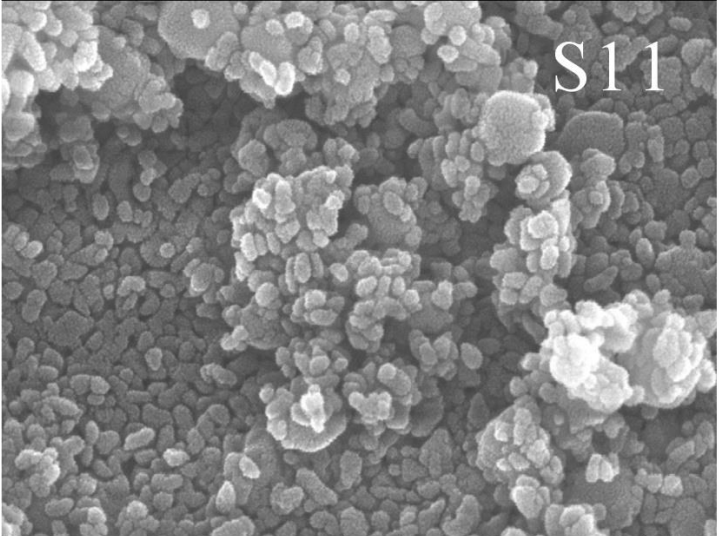

Figure 1: FESEM images of S8, S10 and S11 showing sub-100 nm nanoparticles with a almost spherical morphology, agglomerated in nature.

❖ **Sup.#7:**

**EDX -** Energy Dispersive X-ray Spectroscopy

Elemental compositions were estimated from the 2-D Energy Dispersive X-ray Spectroscopy (EDX) of the different regions of the samples. The EDX scans revealed the presence of only Fe and O peaks and the absence of any other elements other than C. The presence of C is from adventitious C which is generally a feature of any sample and comes from the practical handling of samples (20). The absence of any other impurity elements apart from C ensures the phase purity of the samples. i.e. an undoped or chemically unmodified nature of the samples without the presence of any impurity elements.

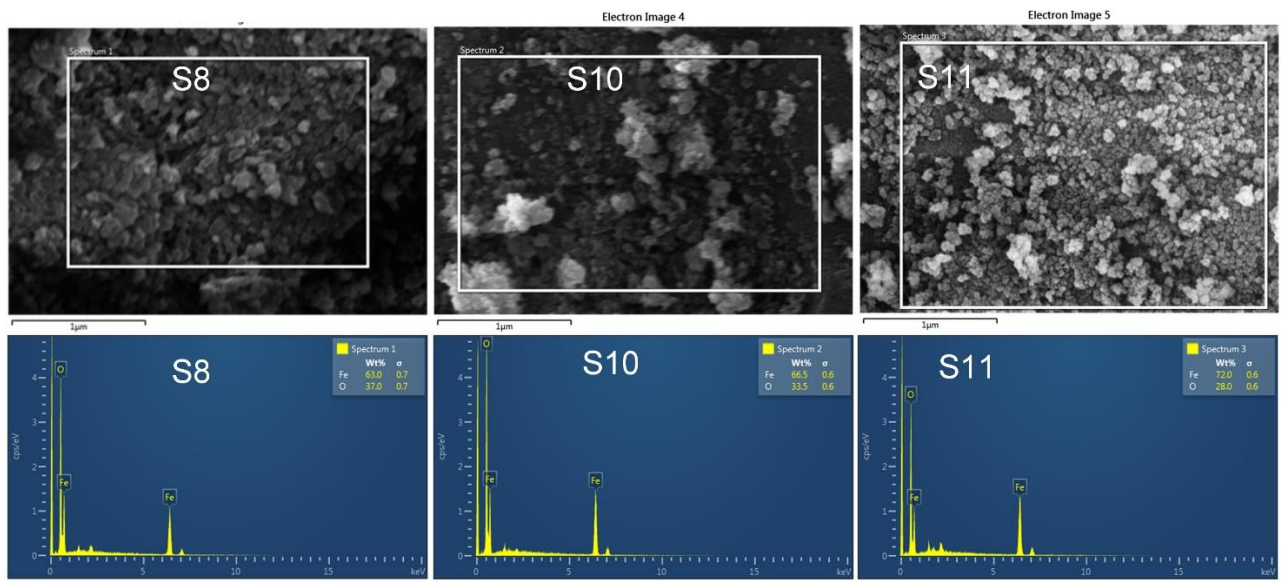

Figure: EDX elemental scans revealing

### ❖ Sup.#8:

**Hydrodynamic Size and Zeta Potential**

1mg/ml of each sample (S8, S10, S11) was prepared in filtered MilliQ and sonicated for 30 minutes at 25°C to form a homogeneous solution. For the PDI and DLS measurements, the homogeneous solution of each sample was transferred to a disposable cuvette (ZEN040).

The colloid-electrolyte interactions are generally investigated from the $\zeta$-potential measurements. The physics of this measurement depends on the interaction between charged particles (electrokinetic phenomena). There will be an attraction between oppositely charged surfaces while similarly charged surfaces are repelled. The electric potential at the slipping plane is termed the $\zeta$-potential. There is a plane that remains attached to the surface and another plane that becomes mobile due to the potential. The interface of these planes is the slipping plane. The zeta potential is measured via the Zetasizer instrument. For the $\zeta$-potential ((21–24), measurements, the samples were transferred to specialized U-shaped zeta cuvettes (DTS1070).

The hydrodynamic size, $d$, $[d = kT/3\pi\eta D$, where $T$ = temperature, $k$ = Boltzmann constant, $D$ = diffusion coefficient] defines the diffusivity of the nano-crystallite within a fluid (25–28). Hence, this parameter is the size of a sphere related to the translational diffusion coefficient of the particle.

On the other hand, a value of PDI > 0.5 indicates polydispersity i.e. the nanoparticles are heterogeneous in solution. Generally, monodisperse particles (PDI < 0.5) are preferred due to lesser variation in their results for various applications.

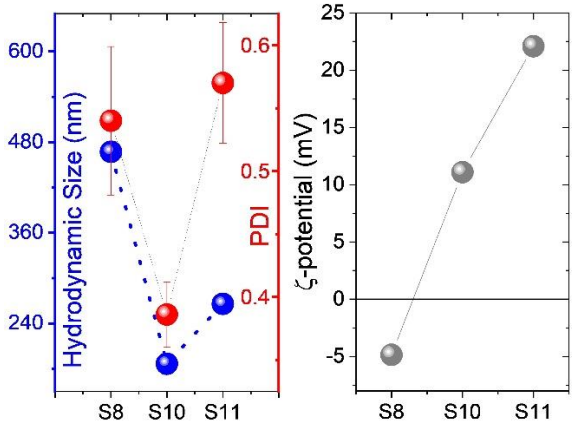

Figure: Hydrothermal Size, Polydispersive Index, and ζ-potential from DLS measurements.

However, it seems that the surface charge of a moderate positive nature could be the reason behind such behavior when a higher positive or negative charge can lead to agglomeration in the fluid. The changes in structure which may lead to such properties need to be analyzed.

## ❖ Sup.#9:

**2.3 X-Ray Diffraction study**

A room temperature X-Ray Diffraction (XRD) study [using a Bruker D2 Phaser x-ray diffractometer] in the range $20°<2\theta<80°$ revealed a crystalline trigonal Hematite ***α***-$Fe_2O_3$ *[space group R-3c]* structure for all the samples as shown in (Figure 2). Rietveld refinement of the XRD data was performed using GSAS software to extract the lattice parameters, strain, and crystallite size. The Crystallographic Index Files (CIF) obtained from the refinement were analyzed in Mercury software and the lattice structure was critically understood.

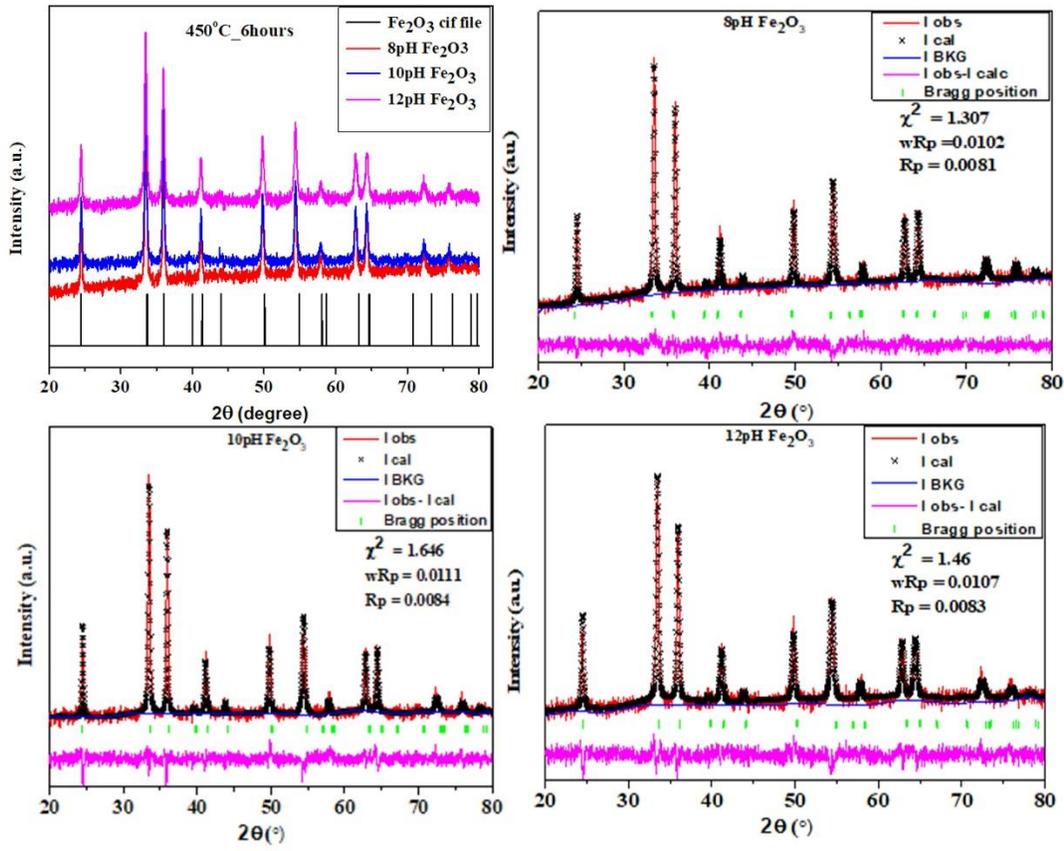

Figure 2: Refined peaks of samples

## ❖ Sup.#10:

**Particle and Crystallite Size**

Rietveld refinement gave the values of fitting parameters $L_x$ and $L_y$. Using the Scherrer formula, the average crystallite size, $D = 18\,000 \cdot (K.\lambda)/\pi L_x$, (where K~1, is a constant) was determined. On the other hand, lattice strain (S) was calculated from $S = \pi L_y/18000$.

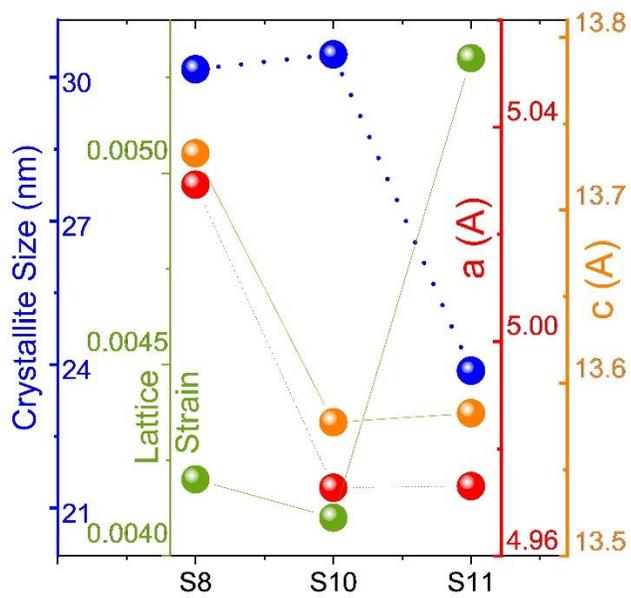

The optimal structural conditions of the S10 sample, as observed from the XRD data reveal an interesting correlation with the ζ-potential and PDI and DLS analysis. The most interesting fact is that the crystallite size of S10 is maximum yet results in the lowest value of the hydrodynamic size, $d$. This can now be well understood from the aspect of lattice strain. Hence, the pH value of ~10 must have some relevance in reducing the strain and hence result in an optimal surface charge which minimizes the hydrodynamic size, $d$. Such a structure can have implications for the particle's interaction with other materials, especially biological species.

❖ **Sup.#11:**

**Bond Lengths and Angles**

A deeper investigation of the structure as depicted in (Figure 3) reveals an interesting Fe-lattice and O-lattice bound to each other in a complex network. The O-lattice consists of layers of O-atoms in a plane with each O-atom being connected to neighboring six O-atoms with four longer bonds $(O-O)_l$ and two shorter bonds $((O-O)_s)$. All these O-atoms have a singular bond length with all the Fe-atoms

(Fe-O). The Fe-atoms are in a staggered plane in between two O-planes. The Fe-Fe bonds with two Fe atoms in the same plane (Fe-Fe)$_p$ are consistently constant for a sample. The subsequent Fe-layers are separated by a characteristic Fe-Fe interplanar bond (Fe-Fe)$_i$, which is shorter than (Fe-Fe)$_p$. The above bond lengths have been studied in detail for these three samples. The values of all these bonds decrease significantly for the S10 and S11 samples as compared to the S8 sample. This proves that the higher alkaline background was effective in reducing all the bond lengths, indicating a better overlap of the electronic orbitals of the constituent atoms. One needs to emphasize that the S10 sample had the least Fe-Fe bond length and was nominally shorter than the S11 sample. These bond lengths have been tabulated in (Table 3). Hence, from all the above studies from XRD, one may conclude that the higher pH helps the materials to ensure a better bond strength of the constituent ions although there seems to be an optimal value of this pH which establishes the best possible bond strength of the same.

The most interesting changes in the O$_s$-Fe-O$_l$ bond angles. This angle was ~89.09º for S8 and 89.1º for S11 but is much smaller, ~87.24º for S10. Hence, due to the pH value of 10, the angular separation of the O-planes containing the O$_s$s with respect to the one containing O$_l$s is reduced. Such a change is a result of the optimal hybridization of electronic orbitals between the Fe 3d and O 2p electrons for a pH value of ~10. This in fact is responsible for the minimum strain in the lattice and must be correlated to the electronic properties. This will be discussed in the next section.

On the other hand, another important observation is the effect of alkalinity on the FeO$_6$ octahedra. Considering the complex network of the hexacoordinated Fe with the quadra-coordinated O atoms, one can easily find that the FeO$_6$ octahedra are distorted. The distortion can be revealed from the two different sets of three Fe-O bonds. The octahedra allow three O atoms of a certain O-plane to be at a shorter Fe-O$_s$ bond with the other three of the adjacent O-plane to be at a longer Fe-O$_l$ bond. The bond lengths, longer and shorter, are the maximum for the S8 sample i.e. 2.135 Å and 1.928 Å. The longer Fe-$O_l$ bond reduces to 2.111Å for both S10 and S11. However, the shorter Fe-$O_s$ bond keeps on reducing to 1.908 Å for S10 and 1.906 Å for S11 as depicted in (Table 3 and Figure 4). Hence, the increasing alkalinity decreases the bond lengths to a considerable extent and especially affects the shorter Fe-$O_s$ bond. Furthermore, this arrangement creates an off-centering of the Fe atom from the center of this octahedron which is more evident from the non-90º O-Fe-O bond angles. The bond angles show a trend of max deformation for the S10 sample, thereby maximizing the off-centering. Hence, it may be possible that the consequence of the continuous reduction in the bond lengths in the $FeO_6$ octahedra, is the off-centering of the Fe-atom. This may be due to the modification of the hybridization of the bonding electrons, due to the ambient environment, i.e. the pH of the solution. These subtle changes thereby bring in modifications to the electronic properties of the materials, and can thereby be correlated to the changes in the bandgap, disorder, and therefore the strain in the lattice.

The strikingly lower value of the $E_U$ value in S10 again refers to a reduced lattice disorder in the sample. This may be a consequence of a reduced lattice strain in the lattice, probably related to the lowest value of $O_l$-Fe-$O_s$ angle, at this optimal pH value. Hence, the involvement of the lattice seems possible in an electron-hole (*e-h*) pair generation. This is a situation of a phonon-assisted *e-h* pair generation, i.e. an indirect bandgap. Hence, from all these studies it is evident that the sample S10 is strikingly different and the changes in the bond properties influence the electronic properties of the materials. As there are reports of $Fe_2O_3$ having a direct as well as an indirect bandgap, it is difficult to comment on the exact nature of bandgap in such samples. A direct bandgap material may also get transformed to an indirect bandgap material due to several reasons related to lattice distortion and disorder. Therefore, a confirmative conclusion on this aspect may not be possible at this stage.

However, from the trends of different parameters one can see a correlation to the structural aspects of the materials. Hence for this study one may accept the indirect bandgap model to be more logical for these studies.

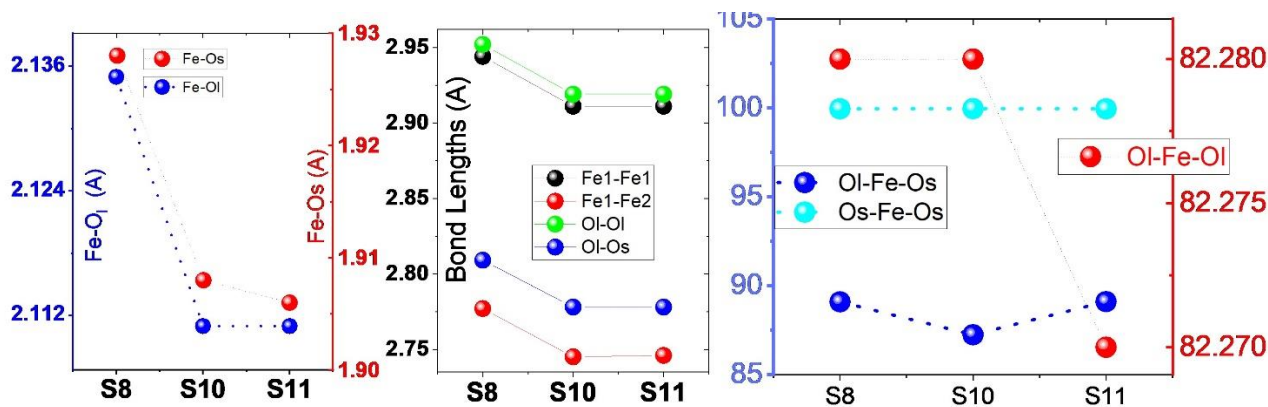

Thereby, it is probable that in such a case, phonon-assisted electron-hole pair generation and assisted ROS generation is possible.

❖ **Sup.#12:**

**Bandgap and Urbach Energy estimation**

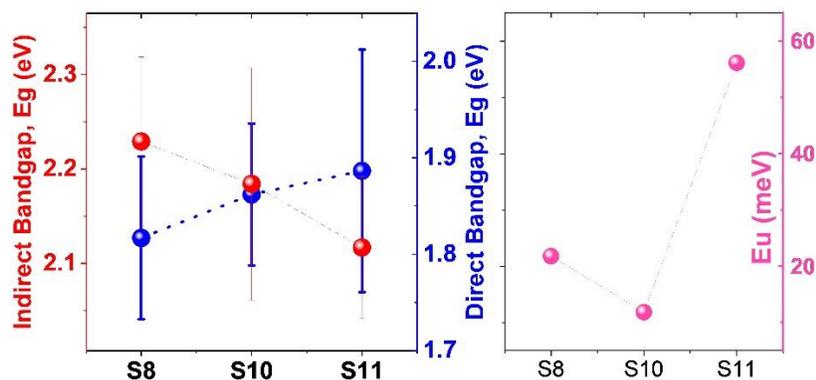

The bandgap ($E_g$) and lattice disorder ($E_U$, Urbach energy) were studied using the optical reflectance, $R$, of the samples using a UV-Vis (Research India Spectrophotometer). The absorption coefficient, $\alpha$ was estimated using the Kubelka-Munk function, $F(R) = A.(1-R)^2/2R$, where, $A$ is the proportionality constant and $E$ is the energy of the incident light. A Tauc plot (29) was used to evaluate the direct and indirect band gaps of the samples.

| Nanoparticle | Average Hydrodynamic size (d.nm) | Average Polydispersity index (PDI) | Average ζ-potential (mV) |
|---|---|---|---|
| S8 | 467.1±43.3 | 0.540±0.059 | -4.85±0.90 |
| S10 | 186.6±2.5 | 0.386±0.026 | 11.10±0.10 |
| S11 | 265.9±15.4 | 0.570±0.048 | 22.10±1.00 |

Table 1: Hydrodynamic size, Polydispersity index and ζ-potential

| Sample | a=b (A) | c (A) | Particle Size [FESEM] (nm) | Crystallite Size [XRD] (nm) | Lattice Strain |
|---|---|---|---|---|---|
| S8 | 5.02925 (0.00013) | 13.73285 (0.00065) | 20~40 | 30.17 | 0.0042 |
| S10 | 4.97272 (0.00014) | 13.57733 (0.0007) | 20~35 | 30.48 | 0.0041 |

| Sample | a=b (A) | c (A) | Particle Size [FESEM] (nm) | Crystallite Size [XRD] (nm) | Lattice Strain |
|---|---|---|---|---|---|
| S11 | 4.97302 (0.00022) | 13.58262 (0.00094) | 10~30 | 23.87 | 0.0053 |

Table 2: Lattice Parameters, Crystalline size and Strain

| Sample | Fe-O$_l$ (A) | Fe-O$_s$ (A) | Fe$_1$-Fe$_1$ (A) | Fe$_1$-Fe$_2$ (A) | O$_l$-O$_l$ (A) | O$_l$-O$_s$ (A) |
|---|---|---|---|---|---|---|
| S8 | 2.135 | 1.928 | 2.944 | 2.777 | 2.952 | 2.809 |
| S10 | 2.111 | 1.908 | 2.911 | 2.745 | 2.919 | 2.778 |
| S11 | 2.111 | 1.906 | 2.911 | 2.746 | 2.919 | 2.778 |

Table 3: Bond lengths

| Sample | O$_l$-Fe-O$_s$ | O$_l$-Fe-O$_l$ | O$_s$-Fe-O$_s$ |
|---|---|---|---|
| S8 | 89.09° | 82.28° | 99.94° |
| S10 | 87.24° | 82.28° | 99.95° |
| S11 | 89.10° | 82.27° | 99.94° |

Table 4: Bond angles

|  | **Direct Eg (eV)** | **Indirect Eg (eV)** | **Urbach energy (meV)** |
|---|---|---|---|
| **S8** | 2.127 ± 0.0863 | 1.917±0.08769 | 21.80±1.14415 |
| **S10** | 2.173 ± 0.0752 | 1.873±0.12022 | 11.86±0.9695 |
| **S11** | 2.198 ± 0.1284 | 1.807±0.07372 | 56.14±1.08007 |

Table 5: Bandgap analysis

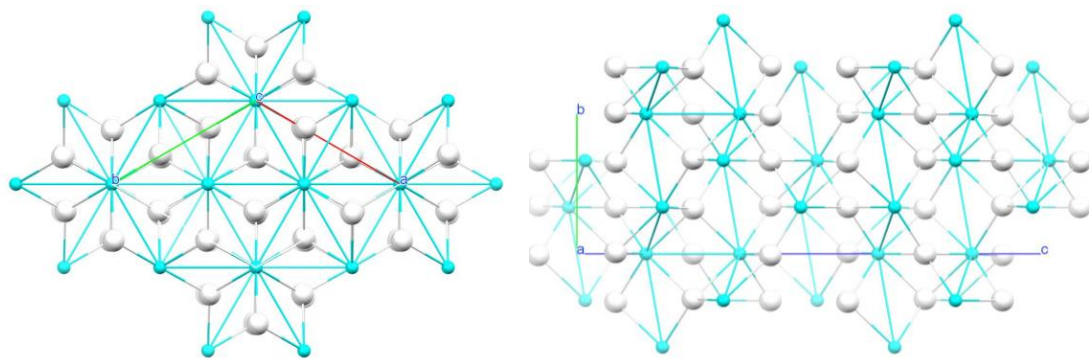

Figure 5: Arrangement of Fe (cyan) and O (white) along the c-axis (left) and b-axis (right). Note the O-plane and the staggered Fe-plane when observed along the b-axis.

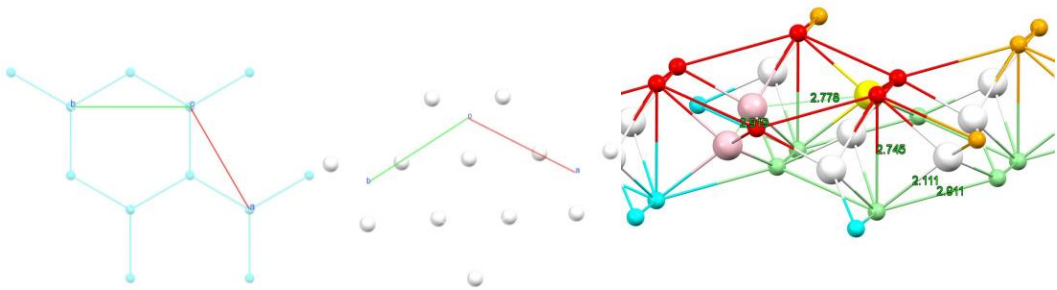

Figure 6: Fe-Fe network and oxygen network

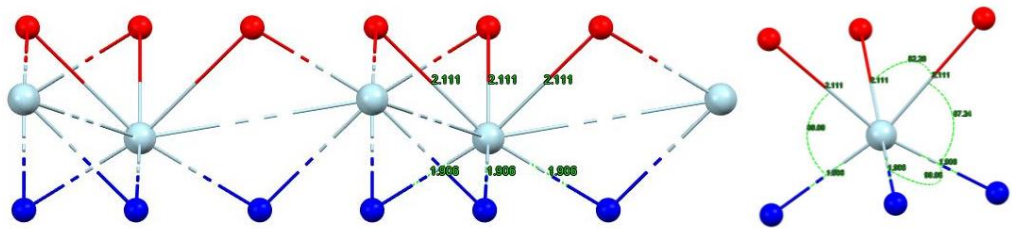

Figure 7: Planar depiction of $Fe_2O_3$ and $FeO_6$ octahedra

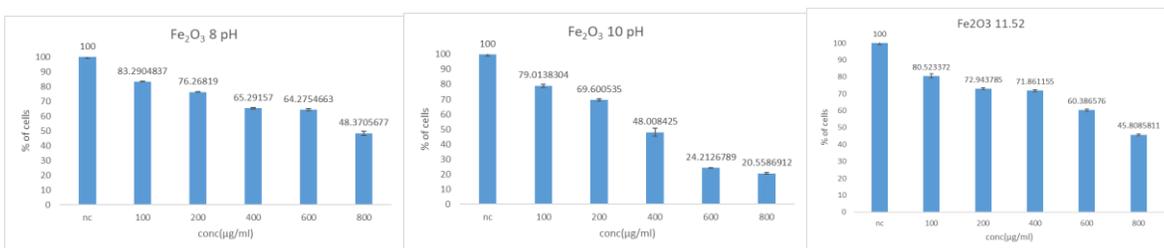

Figure 8: Minimum inhibitory concentration (MIC) of the samples

1. Vu XH, Duong TTT, Pham TTH, Trinh DK, Nguyen XH, Dang VS. Synthesis and study of silver nanoparticles for antibacterial activity against *Escherichia coli* and *Staphylococcus aureus*. Advances in Natural Sciences: Nanoscience and Nanotechnology. 2018 Jun 8;9(2):025019.

2. Mendes CR, Dilarri G, Forsan CF, Sapata V de MR, Lopes PRM, de Moraes PB, et al. Antibacterial action and target mechanisms of zinc oxide nanoparticles against bacterial pathogens. Sci Rep. 2022 Feb 16;12(1):2658.
3. Shamaila S, Zafar N, Riaz S, Sharif R, Nazir J, Naseem S. Gold Nanoparticles: An Efficient Antimicrobial Agent against Enteric Bacterial Human Pathogen. Nanomaterials. 2016 Apr 14;6(4):71.
4. Ali M, Ijaz M, Ikram M, Ul-Hamid A, Avais M, Anjum AA. Biogenic Synthesis, Characterization and Antibacterial Potential Evaluation of Copper Oxide Nanoparticles Against Escherichia coli. Nanoscale Res Lett. 2021 Dec 20;16(1):148.
5. Singh V, Yadav SS, Chauhan V, Shukla S, Vishnolia KK. Applications of Nanoparticles in Various Fields. In 2021. p. 221–36.
6. Kelly KL, Coronado E, Zhao LL, Schatz GC. The Optical Properties of Metal Nanoparticles: The Influence of Size, Shape, and Dielectric Environment. J Phys Chem B. 2003 Jan 1;107(3):668–77.
7. Nazanin Abbaspour, Richard Hurrell, Roya Kelishadi. Review on iron and its importance for human health. Journal of Research in Medical Sciences. 2014 Feb;
8. Iqbal Ahmed, Kamisah Kormin, Rizwan Rajput, Muhammed H Albeirutty. The Importance of Iron oxides in Natural Environment and Significance of its Nanoparticles Application. In: Ahmed I, Kormin K, Rajput R, Albeirutty MH, Rehan ZA, Zeb J, editors. 2018. p. 218–57.
9. Wollschläger J. Reactive Molecular Beam Epitaxy of Iron Oxide Films: Strain, Order, and Interface Properties. In: Encyclopedia of Interfacial Chemistry. Elsevier; 2018. p. 284–96.
10. Vihodceva S, Šutka A, Sihtmäe M, Rosenberg M, Otsus M, Kurvet I, et al. Antibacterial Activity of Positively and Negatively Charged Hematite (α-Fe2O3) Nanoparticles to Escherichia coli, Staphylococcus aureus and Vibrio fischeri. Nanomaterials. 2021 Mar 8;11(3):652.
11. Khatami M, Aflatoonian MR, Azizi H, Mosazade F, Hooshmand A, Lima Nobre MA, et al. Evaluation of Antibacterial Activity of Iron Oxide Nanoparticles Against Escherichia coli. International Journal of Basic Science in Medicine. 2017 Dec 31;2(4):166–9.
12. Sankadiya S, Oswal N, Jain P, Gupta N. Synthesis and characterization of Fe2O3 nanoparticles by simple precipitation method. In 2016. p. 020064.
13. Majid Farahmandjou. Synthesis and characterization of α-Fe2O3 nanoparticles by simple co-precipitation method. Research Letters in Physical Chemistry. 2015 Sep;
14. Szterner P, Biernat M. The Synthesis of Hydroxyapatite by Hydrothermal Process with Calcium Lactate Pentahydrate: The Effect of Reagent Concentrations, pH, Temperature, and Pressure. Bioinorg Chem Appl. 2022 Mar 25;2022:1–13.
15. Wilhelm MJ, Sharifian Gh. M, Wu T, Li Y, Chang CM, Ma J, et al. Determination of bacterial surface charge density via saturation of adsorbed ions. Biophys J. 2021 Jun;120(12):2461–70.
16. Ismail RA, Sulaiman GM, Abdulrahman SA, Marzoog TR. Antibacterial activity of magnetic iron oxide nanoparticles synthesized by laser ablation in liquid. Materials Science and Engineering: C. 2015 Aug;53:286–97.
17. Sonohara R, Muramatsu N, Ohshima H, Kondo T. Difference in surface properties between Escherichia coli and Staphylococcus aureus as revealed by electrophoretic mobility measurements. Biophys Chem. 1995 Aug;55(3):273–7.

18. Zhao X, Drlica K. Reactive oxygen species and the bacterial response to lethal stress. Curr Opin Microbiol. 2014 Oct;21:1–6.
19. V. LP, Vijayaraghavan R. Chemical manipulation of oxygen vacancy and antibacterial activity in ZnO. Materials Science and Engineering: C. 2017 Aug;77:1027–34.
20. Greczynski G, Hultman L. The same chemical state of carbon gives rise to two peaks in X-ray photoelectron spectroscopy. Sci Rep. 2021 May 27;11(1):11195.
21. Arakha M, Saleem M, Mallick BC, Jha S. The effects of interfacial potential on antimicrobial propensity of ZnO nanoparticle. Sci Rep. 2015 Apr 15;5(1):9578.
22. Chang SH, Lin HTV, Wu GJ, Tsai GJ. pH Effects on solubility, zeta potential, and correlation between antibacterial activity and molecular weight of chitosan. Carbohydr Polym. 2015 Dec;134:74–81.
23. Chen LC, Kung SK, Chen HH, Lin SB. Evaluation of zeta potential difference as an indicator for antibacterial strength of low molecular weight chitosan. Carbohydr Polym. 2010 Oct;82(3):913–9.
24. Ferreyra Maillard APV, Espeche JC, Maturana P, Cutro AC, Hollmann A. Zeta potential beyond materials science: Applications to bacterial systems and to the development of novel antimicrobials. Biochimica et Biophysica Acta (BBA) - Biomembranes. 2021 Jun;1863(6):183597.
25. Azam A. Size-dependent antimicrobial properties of CuO nanoparticles against Gram-positive and -negative bacterial strains. Int J Nanomedicine. 2012 Jul;3527.
26. Heravi MEM. Effects of Hydrodynamic Diameter of Nanoparticles on Antibacterial Activity and Durability of Ag-treated Cotton Fabrics. Fibers and Polymers. 2020 Jun 23;21(6):1173–9.
27. Dadi R, Azouani R, Traore M, Mielcarek C, Kanaev A. Antibacterial activity of ZnO and CuO nanoparticles against gram-positive and gram-negative strains. Materials Science and Engineering: C. 2019 Nov;104:109968.
28. Azam A, Ahmed, Oves, Khan, Habib, Memic A. Antimicrobial activity of metal oxide nanoparticles against Gram-positive and Gram-negative bacteria: a comparative study. Int J Nanomedicine. 2012 Dec;6003.
29. Landi S, Segundo IR, Freitas E, Vasilevskiy M, Carneiro J, Tavares CJ. Use and misuse of the Kubelka-Munk function to obtain the band gap energy from diffuse reflectance measurements. Solid State Commun. 2022 Jan;341:114573.